\documentclass[sma4]{snow2e}
\usepackage{graphics}

\begin{document}

\title{Measuring Chiral Parameters 
	in the Strongly Interacting $W$ System\\ 
	at a Linear Collider\thanks{Work submitted to the 1996
		Snowmass Conference in collaboration
		with E.~Boos, A.~Pukhov (Moscow) and 
		H.-J.~He, P.M.~Zerwas (DESY).}}

\author{W.\ Kilian\\ {\it DESY, Theory Group, 22603 Hamburg, Germany}}

\maketitle

%% Get rid of page numbering
%\thispagestyle{empty}\pagestyle{empty}

\begin{abstract} 
In the absence of a Higgs particle vector boson scattering amplitudes
are generally described by an electroweak chiral Lagrangian below the
resonance region.  For a Linear Collider with CMS energy
$\sqrt{s}=1.6\ {\rm TeV}$ and an integrated luminosity of $200\ {\rm
fb}^{-1}$ we estimate the sensitivity on the chiral parameters
$\alpha_4$ and $\alpha_5$.  We consider the processes $e^+e^-\to
W^+W^-\bar\nu\nu$ and $e^+e^- \to ZZ\bar\nu\nu$, performing a
complete calculation which includes all relevant Feynman diagrams at
tree level, without relying on the Equivalence Theorem or the
Effective $W$ Approximation.  The dominant backgrounds and $W/Z$
misidentification probabilities are accounted for.
\end{abstract}

\section{Introduction}
Without the cancellations induced by a Higgs resonance, the scattering
amplitudes of massive vector bosons grow with rising energy,
saturating the unitarity bounds in the TeV region~\cite{Uni}.  Thus
there is a strongly interacting domain which lies within the reach of
the next generation of collider experiments.  One usually expects new
resonances which manifest themselves as peaks in the invariant mass
distribution of massive vector boson pairs $VV$ in reactions which
contain the nearly on-shell scattering $V'V'\to VV$ as a subprocess.

As Barger et al.\ have shown~\cite{BCHP}, with a suitable set of
kinematical cuts different resonance models (in particular, a $1 {\rm
TeV}$ scalar and a $1\ {\rm TeV}$ vector) can clearly be distinguished
by analyzing the two modes $e^+e^-\to W^+W^-\bar\nu\nu$ and $e^+e^-\to
ZZ\bar\nu\nu$ at a Linear Collider with $1.5\ {\rm TeV}$ CMS energy.
A number of similar analyses for hadron, $e^-e^-$, and muon collisions
have also been performed~\cite{LHC,eminus,muon}.

This result encourages one to consider the same processes in the more
difficult case when resonances do not exist or are out of reach of the
particular experiment.  In the following we present results for the
sensitivity on details of the strong interactions as a two-parameter
analysis, carried out in the framework of a complete tree-level
calculation.

\section{Chiral Lagrangian}
Below a suspected resonance region the electroweak theory is properly
parameterized in terms of a gauged chiral Lagrangian which
incorporates the spontaneous breaking of the electroweak symmetry.
This Lagrangian induces a low-energy approximation of scattering
amplitudes organized in powers of the energy~\cite{ChPT}.

Models for strong vector boson scattering are usually embedded in
Standard Model calculations via the Equivalence Theorem~\cite{ET}
and/or the Effective $W$ Approximation~\cite{EWA}.  However, they are
not needed for our purpose, since the very nature of chiral Lagrangians
as effective low-energy theories allows a complete calculation without
approximations.  For an accurate estimate of the sensitivity the
correct treatment of transversally polarized vector bosons and
interference effects is essential, and the full kinematics of the
process must be known in order to sensibly apply cuts necessary to
isolate the signal.

In our study we made use of the automated calculation package
CompHEP~\cite{CompHEP}.  For technical reasons, the chiral Lagrangian
has been implemented in 'tHooft-Feynman gauge:
\begin{equation}
  {\cal L} = {\cal L}_{\rm G} + {\cal L}_{\rm GF} + {\cal L}_{\rm FP}
		+ {\cal L}_e
		+ {\cal L}_0 + {\cal L}_4 + {\cal L}_5
\end{equation}
where
\begin{eqnarray}
  {\cal L}_{\rm G} &=& -\textstyle\frac18{\rm tr}[W_{\mu\nu}^2]
		- \textstyle\frac14 B_{\mu\nu}^2\\
 {\cal L}_{\rm GF} &=& 
		- \textstyle\frac12 \left(\partial^\mu W_\mu^a
		+ i\frac{gv^2}{4}{\rm tr}[U\tau^a]\right)^2\nonumber\\
	&&
		-{} \textstyle\frac12 \left(\partial^\mu B_\mu
		- i\frac{g'v^2}{4}{\rm tr}[U\tau^3]\right)^2\\
  {\cal L}_e	&=& \bar e_{\rm L}iD\!\!\!\!/\,\,e_{\rm R} 
		+ \bar\nu_{\rm L}iD\!\!\!\!/\,\,e_{\rm R} 
		+ \mbox{h.c.}\\
  {\cal L}_0	&=& \textstyle\frac{v^2}{4}{\rm tr}
		[D_\mu U^\dagger D^\mu U]\\
  {\cal L}_4	&=& \alpha_4\,{\rm tr}[V_\mu V_\nu]^2 \\
  {\cal L}_5	&=& \alpha_5\,{\rm tr}[V_\mu V^\mu]^2
\end{eqnarray}
with the definitions
\begin{eqnarray}
  U &=& \exp(-iw^a\tau^a/v) \\
  V_\mu &=& U^\dagger D_\mu U
\end{eqnarray}

\section{Parameters}
To leading order the chiral expansion contains two independent
parameters which give rise to $W$ and $Z$ masses.  The fact that they
are related, \emph{i.e.}, the $\Delta\rho$ (or $\Delta T$) parameter
is close to zero, suggests that the new strong interactions respect a
custodial $SU_2^L\times SU_2^R$ symmetry~\cite{SU2c}, spontaneously
broken to the diagonal $SU_2$.

In next-to-leading order there are eleven CP-even chiral parameters.
Two of them correspond to the $S$ and $U$ parameters~\cite{STU}.  Four
additional parameters describe the couplings of three gauge bosons.
They can be determined, e.g., at $e^+e^-$ colliders by analyzing $W$
boson pair production~\cite{TGV}.  In our study we assume that these
parameters are known with sufficient accuracy.  For simplicity, we set
them to zero.

The remaining five parameters are visible only in vector boson
scattering.  If we assume manifest custodial symmetry, only two
independent parameters $\alpha_4$ and $\alpha_5$ remain.  They can be
determined by measuring the total cross section of vector boson
scattering in two different channels.

In the present study we consider the two channels $W^+W^-\to W^+W^-$
and $W^+W^-\to ZZ$ which are realized at a Linear Collider in the
processes $e^+e^-\to W^+W^-\bar\nu\nu$ and $e^+e^-\to ZZ\bar\nu\nu$.
In the limit of vanishing gauge couplings the amplitudes for the two
subprocesses are related:
\begin{eqnarray}
  a(W^+_L W^-_L\to Z_LZ_L) &=& A(s,t,u)\\
  a(W^+_L W^-_L\to W^+_LW^-_L) &=& A(s,t,u) + A(t,s,u)
\end{eqnarray}
where
\begin{equation}
  A(s,t,u) = \frac{s}{v^2} + \alpha_4\frac{4(t^2+u^2)}{v^4} 
		+ \alpha_5\frac{8s^2}{v^4}
\end{equation}
with $v=246\ {\rm GeV}$.  These relations hold only for the
longitudinal polarization modes.  Although in the present study all
modes are included, they lead us to expect an increase in the
rate for both processes with positive $\alpha_4$ and $\alpha_5$.
Negative values tend to reduce the rate as long as the leading term is
not compensated.

\section{Calculation}
Using the above Lagrangian, the full squared matrix elements for the
processes $e^+e^-\to W^+W^-\bar\nu\nu$ and $e^+e^-\to ZZ\bar\nu\nu$
have been analytically calculated and numerically integrated at
$\sqrt{s}=1600\ {\rm GeV}$ (omitting $Z$ decay diagrams, see below).
The backgrounds $e^+e^-\to W^+W^- e^+e^-$ and $e^+e^-\to W^\pm Z
e^\mp\nu$ are relevant if the electrons escape undetected through the
beampipe.  In that region they receive their dominant contribution
through $\gamma\gamma$, $\gamma Z$ and $\gamma W$ fusion which has
been calculated within the Weizs\"acker-Williams
approximation~\cite{EPA}.

A set of optimized cuts to isolate various strongly interacting $W$
signals has been derived in~\cite{BCHP}.  It turns out that similar
cuts are appropriate in our case:
\begin{center}
  $|\cos\theta(W)|<0.8$ \nonumber\\
  $150\ {\rm GeV}<p_T(W)$ \nonumber\\
  $50\ {\rm GeV} < p_T(WW) < 300\ {\rm GeV}$ \nonumber\\
  $200\ {\rm GeV} < M_{\rm inv}(\bar\nu\nu)$ \nonumber\\
  $700\ {\rm GeV} < M_{\rm inv}(WW) < 1200\ {\rm GeV}$
\end{center}
The lower bound on $p_T(WW)$ is necessary because of the large $W^+W^-
e^+e^-$ background which is concentrated at low $p_T$ if both
electrons disappear into the beampipe.  We have assumed an effective
opening angle of $10$ degrees.  The cut on the $\bar\nu\nu$ invariant
mass removes events where the neutrinos originate from $Z$ decay,
together with other backgrounds~\cite{BCHP}.  For the $ZZ$ final state
the same cuts are applied, except for $p_T^{\rm min}(ZZ)$ which can be
reduced to $30\ {\rm GeV}$.

The restriction to a window in $M_{\rm inv}(WW)$ between $700$ and
$1200\ {\rm GeV}$ keeps us below the region where (apparent) unitarity
violation becomes an issue.  Furthermore, it fixes the scale of the
measured $\alpha$ values, which in reality are running parameters, at
about $1\ {\rm TeV}$.  In any case, including lower or higher
invariant mass values does not significantly improve the results.

For the analysis we use hadronic decays of the $W^+W^-$ pair and
hadronic as well as $e^+e^-$ and $\mu^+\mu^-$ decays of the $ZZ$ pair.
In addition, we have considered $WW\to jj\ell\nu$ decay modes which
are more difficult because of the additional neutrino in the final
state.  We find that with appropriately modified cuts the backgrounds
can be dealt with also in that case, although the resulting
sensitivity is lower than for hadronic decays.  In the following
results the leptonic $W$ decay modes are not included.

We adopt the dijet reconstruction efficiencies and misidentification
probabilities that have been estimated in~\cite{BCHP}.  Thus we assume
that a true $W$ ($Z$) dijet will be identified as follows:
\begin{eqnarray}
  W &\to& 85\%\;W,\ 10\%\;Z,\ 5\%\;\mbox{reject}\\
  Z &\to& 22\%\;W,\ 74\%\;Z,\ 4\%\;\mbox{reject}
\end{eqnarray}
With $b$ tagging the $Z\to W$ misidentification probability could be
further reduced, improving the efficiency in the $ZZ$ channel.

Including the branching ratios and a factor $2$ for the $WZ$
background, we have the overall efficiencies
\begin{eqnarray}\label{eps}
  \epsilon(WW) &=& 34\% \nonumber\\
  \epsilon(ZZ) &=& 34\% \\
  \epsilon(WZ) &=& 18\%\;\mbox{id.~as $WW$},\ 8\%\;\mbox{as $ZZ$}\nonumber
\end{eqnarray}

\begin{figure}[htb]
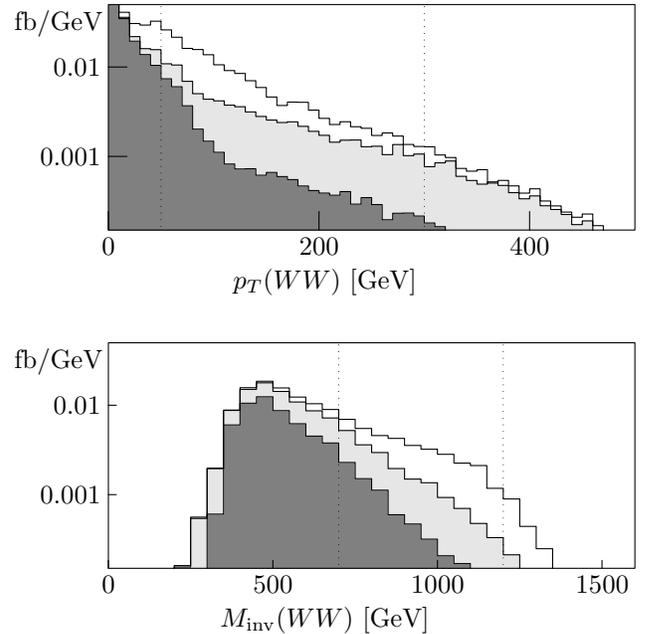

\unitlength1mm
\leavevmode
\begin{center}
\begin{picture}(80,80)
\put(10,50){\includegraphics{ptWW.1}}
\put(10,5){\includegraphics{mWW.1}}
\end{picture}
\end{center}
\caption{Differential distributions in $p_T$ and $M_{\rm inv}$ of the
$W$ pair (after cuts).  The dark area shows the background from $WWee$
and $WZe\nu$ final states; the light area is the rate after the signal
process $e^+e^-\to W^+W^-\bar\nu\nu$ with $\alpha_4=\alpha_5=0$ has
been added; the upper curve denotes the corresponding distribution for
$\alpha_4=0$, $\alpha_5=0.005$.  The $WW$ reconstruction efficiency
has not been included.}
\label{WWplots} 
\end{figure}

\section{Results}
The simulations have been carried out for a number of different values
of the two parameters $\alpha_4$ and $\alpha_5$, such that a
two-parameter analysis was possible for all observables.
Fig.~\ref{WWplots} shows the differential distributions in the
transverse momentum and invariant mass of the $WW$ pair for $e^+e^-\to
W^+W^-\bar\nu\nu$ including backgrounds after all cuts have been
applied.  The shown signal distribution is similar in shape to a
broad scalar (Higgs) resonance; however, the total signal rate is
smaller.

Both channels are enhanced by positive values of the two parameters,
the $ZZ$ channel being less sensitive to $\alpha_4$ than the $WW$
channel.  With actual data at hand one would perform a
maximum-likelihood fit to the various differential distributions.  In
our analysis, however, we only use the total cross sections after
cuts.  For $\alpha_4=\alpha_5=0$ we find $80$ $WW$ and $67$ $ZZ$
events if $200\ {\rm fb}^{-1}$ of integrated luminosity with
unpolarized beams and the efficiencies~(\ref{eps}) are assumed.

\begin{figure}[hbt]
\unitlength 1mm
\leavevmode
\begin{center}
\begin{picture}(80,74)
\put(10,3){\includegraphics{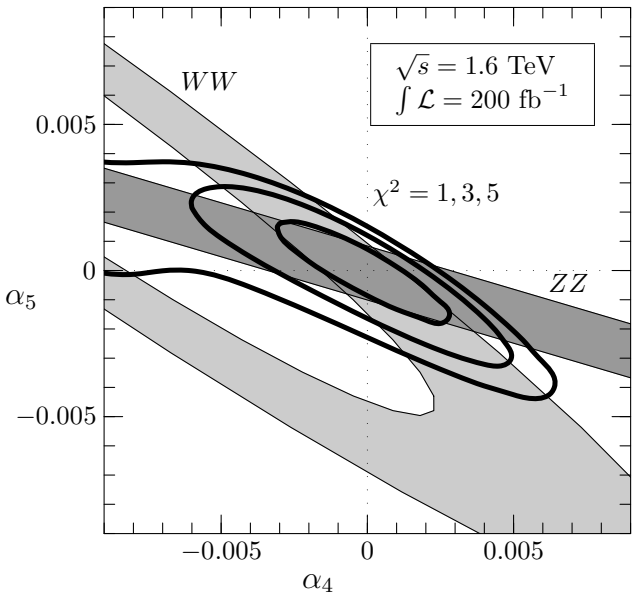}}
\end{picture}
\end{center}
\caption{Exclusion limits for unpolarized beams.  The shaded bands
display the $\pm 1\sigma$ limits resulting from either one of the two
channels; the lines show the combined limits at the $\chi^2=1,3,5$ level.
[For Gaussian distributions, this corresponds to a $39\%$, $78\%$,
$92\%$ confidence level, respectively.]}
\label{contour}
\end{figure}

In Fig.~\ref{contour} we show the $\pm 1\sigma$ bands resulting from
the individual channels as well as the two-parameter confidence region
centered at $(0,0)$ in the $\alpha_4$-$\alpha_5$ plane.  The total
event rate allows for a second solution centered roughly at
$(-0.017,0.005)$ which corresponds to the case where the
next-to-leading contributions in the chiral expansions are of opposite
sign and cancel the leading-order term.  This might be considered as
unphysical; in any case, this part of parameter space could be ruled
out by performing a fit to the differential distributions or by
considering other channels such as $WZ$ [possibly including results
from the LHC].

Since in both channels the signal part is generated only by the
combination of left-handed electrons and right-handed positrons,
polarizing the incident beams enhances the sensitivity of the
experiments.  Assuming $90\%$ electron and $60\%$ positron
polarization, the signal rate increases by a factor $3$.  For
the $WZ$ background the enhancement is $1.75$, whereas the
$W^+W^-e^+e^-$ background remains unchanged.  We now find $182$ $WW$
and $193$ $ZZ$ events.  

\begin{figure}[hbt]
\unitlength 1mm
\leavevmode
\begin{center}
\begin{picture}(80,74)
\put(10,3){\includegraphics{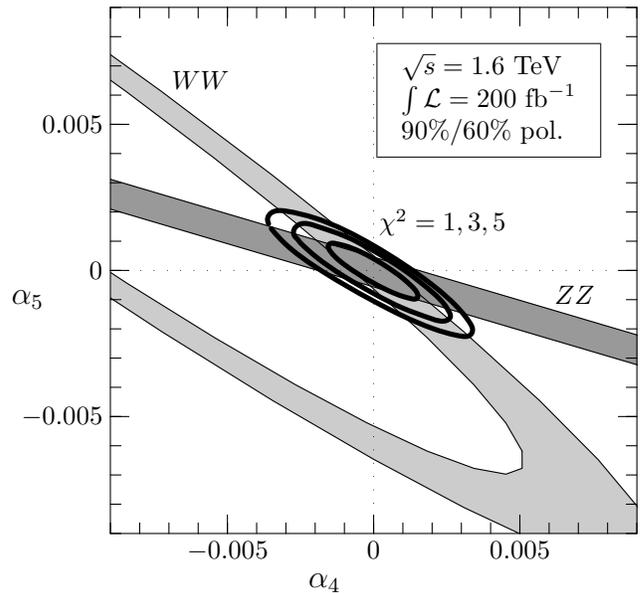}}
\end{picture}
\end{center}
\caption{Exclusion limits for polarized beams.}
\label{contour-pol}
\end{figure}

Here we have not taken into account that part of the intrinsic
background to $e^+e^-\to W^+W^-(ZZ)\bar\nu\nu$ is not due to $WW$
fusion diagrams and will therefore not be enhanced, and that the cuts
could be further relaxed in the polarized case.  Thus the actual
sensitivity will be improved even more.

\section{Summary}
As our analysis shows, a Linear Collider is able to probe the chiral
parameters $\alpha_4$ and $\alpha_5$ down to a level of $10^{-3}$
which is well in the region where the actual values are expected by
dimensional analysis.  Full energy ($\sqrt{s}=1.6\ {\rm TeV}$) and
full luminosity ($200\ {\rm fb}^{-1}$) is needed to achieve that goal.
Electron and positron beam polarization both improve the sensitivity.
With several years of running time a precision measurement of chiral
parameters seems to be a realistic perspective, rendering a meaningful
test of strongly interacting models even in the pessimistic case where
no resonances can be observed directly.

%%%%% References
%

\end{document}